# Comment on "The Sun is less active than other solar-like stars"

**Travis S. Metcalfe[1] and Jennifer van Saders[2]**
[1] Space Science Institute, Boulder CO 80301, USA.
[2] Institute for Astronomy, University of Hawai'i, Honolulu HI 96822, USA.

**Reinhold et al. (Reports, 1 May 2020, p. 518) provided two possible interpretations of measurements showing that the Sun is less active than other solar-like stars. We argue that one of those interpretations anticipates the observed differences between the properties of their two stellar samples. This suggests that solar-like stars become permanently less variable beyond a specific evolutionary phase.**

Reinhold *et al.* (*1*) use observations from the *Kepler* space telescope and the *Gaia* mission to identify a sample of 369 "solar-like" stars that resemble the Sun in terms of surface temperature and composition, rotation period, and age. Using similar selection criteria, they also identify a sample of 2529 "pseudo-solar" stars for which a rotation period could not be determined. For each star, they measure a variability range $R_{var}$ from four years of *Kepler* observations, which they compare to analogous measurements of solar variability.

For the one-eighth of the composite sample with measured rotation (periodic stars), the targets span the full range of variability, from sun-like to five times more variable. For the larger sample with unknown rotation (non-periodic stars), the targets typically show much lower variability, almost never exceeding twice the maximum variability observed in the Sun. Based on these measurements, Reinhold *et al.* conclude that either there are unidentified differences between the two samples of stars, or the observed distribution represents the full range of variability that the Sun can potentially exhibit.

From the perspective of stellar evolution theory, there are substantial differences between the observed properties of the periodic and non-periodic stars. The periodic stars are typically cooler and slightly more metal-rich, while the non-periodic stars are often hotter and metal-poor compared to the Sun (Fig. 1). These differences produce a systematic bias toward deeper convective envelopes in periodic stars, and shallower convective envelopes in non-periodic stars (*2*). Along with rotation, the properties of convective envelopes have a strong influence on the variability of solar-like stars (*3, 4*).

The observed differences between these two samples suggest an evolutionary connection, with periodic stars gradually becoming non-periodic stars. Observations of solar-like stars with asteroseismic ages revealed that rotation and magnetic activity decouple when stars reach a critical value of the Rossby number ($Ro$), the rotation period normalized by the convective turnover time (*5, 6*). Stars like the Sun reach this critical value of $Ro$ near the solar age (*7*), while stars with shallower convective envelopes reach it earlier and those with deeper convective envelopes reach it later (*5, 8*). The decoupling coincides with a change in the properties of stellar magnetic cycles (*9*), possibly triggered by the disruption of differential rotation (*6, 10*), and it appears to be associated with the disappearance of a global dipole magnetic field (*11, 12*).

If solar-like stars near the age of the Sun transition from higher to lower levels of variability at a critical value of $Ro$, then the composite sample of Reinhold *et al.* should contain stars with a range of values. Consequently, the detection of rotation and the observed rotation periods should be biased toward stars with lower $Ro$ and away from those with higher $Ro$. These observational signatures of the suggested magnetic transition in solar-like stars are apparent in the Reinhold *et al.* sample (Fig. 2). The alternative interpretation, that stars like the Sun may occasionally exhibit much higher levels of activity, would not be expected to show a dependence on $Ro$.

Reinhold *et al.* acknowledged the influence of stellar properties on the observed variability, and attempted to remove these dependencies using a multivariate linear regression model. In their Supplementary Materials (Figs. S8 and S9) they quantify the change in the variability range $R_{var}$ with temperature, composition, and rotation period. For each star in the periodic sample, they correct the values of $R_{var}$ to reflect the expected value if the stellar properties were identical to those of the Sun. Due to the large uncertainties in temperature (150-200 K) and composition (0.15-0.30 dex), these corrections are approximate (*13*). For the non-periodic sample, $R_{var}$ did not apparently depend on temperature or composition, so no correction was applied.

The corrected variability range does not completely account for differences between the ages of stars in the periodic sample, and fails to consider possible evolutionary effects within the composite sample. If the variability is expected to change near the age of the Sun, then the selection of stars between isochrones at 4 and 5 Gyr will span the transition. Although a correction for rotation should remove much of the dependence on evolutionary phase for stars below the critical value of $Ro$, it will undercorrect more evolved stars (*14*).

Solar-like stars become more luminous as they evolve, so the *Gaia* absolute magnitude ($M_G$) can be used as a proxy for the evolutionary phase. Reinhold *et al.* did not include $M_G$ in their multivariate analysis, but the corrected values of $R_{var}$ for the periodic sample still depend on $M_G$, revealing an uncorrected age dependence. The relative fractions of periodic and non-periodic stars in each range of $M_G$ reinforces this conclusion, with non-periodic stars accounting for 97% of the most luminous targets, and the periodic fraction growing larger for the least evolved stars. Accounting for this residual age dependence may not significantly change the variability distribution of the composite sample, but it will likely reveal that periodic stars are typically less evolved than the Sun, while non-periodic stars are more evolved.

Considering that 87% of the Reinhold *et al.* sample exhibits low levels of variability like the Sun, their results may be the best evidence yet that our star is already transitioning to a magnetically inactive future.

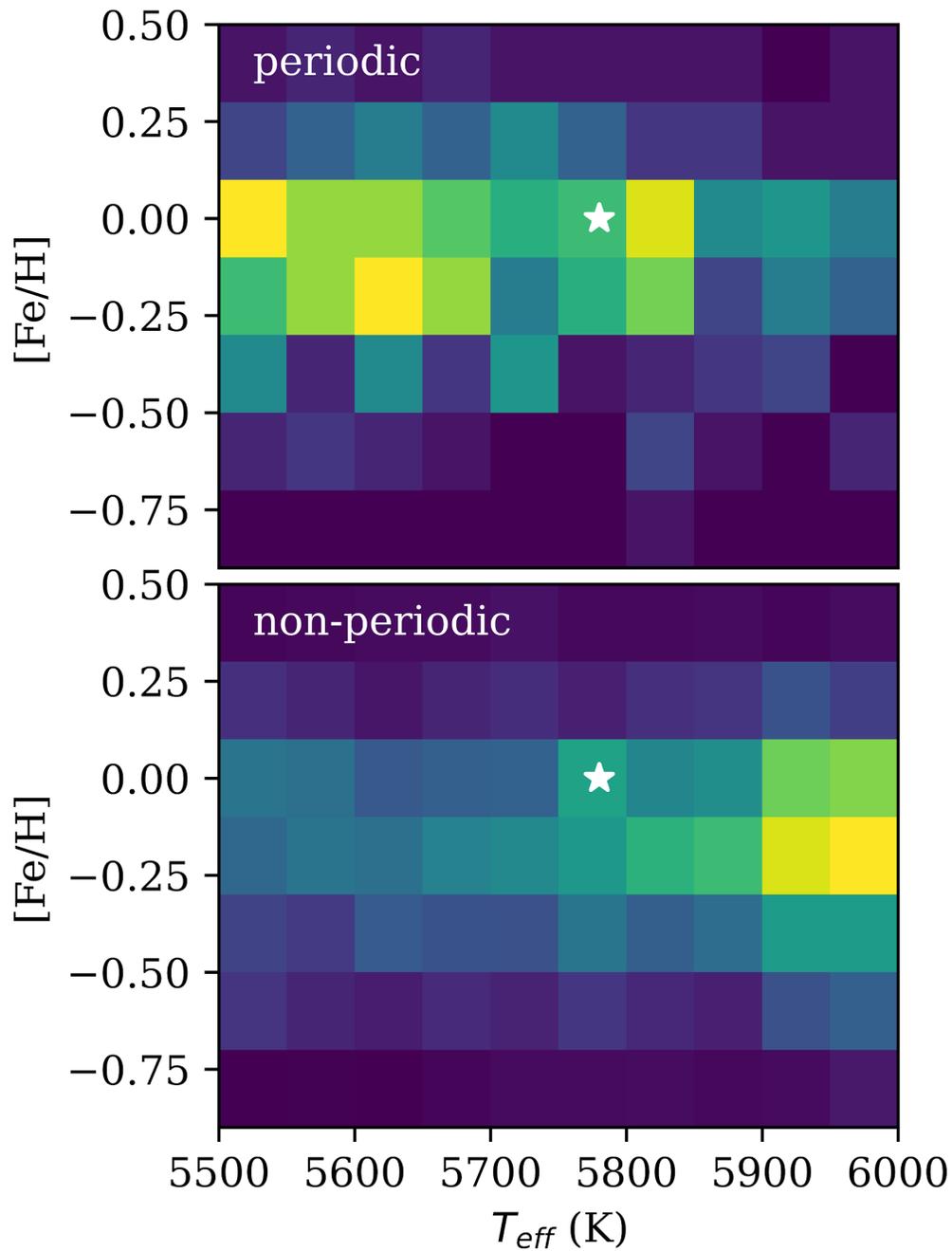

**Fig. 1. Distributions of surface temperature and composition for the periodic (A) and non-periodic (B) samples.** The differences between the two samples produce a systematic bias in the properties of their convective envelopes, which have a strong influence on their variability. The Sun is indicated with a white star.

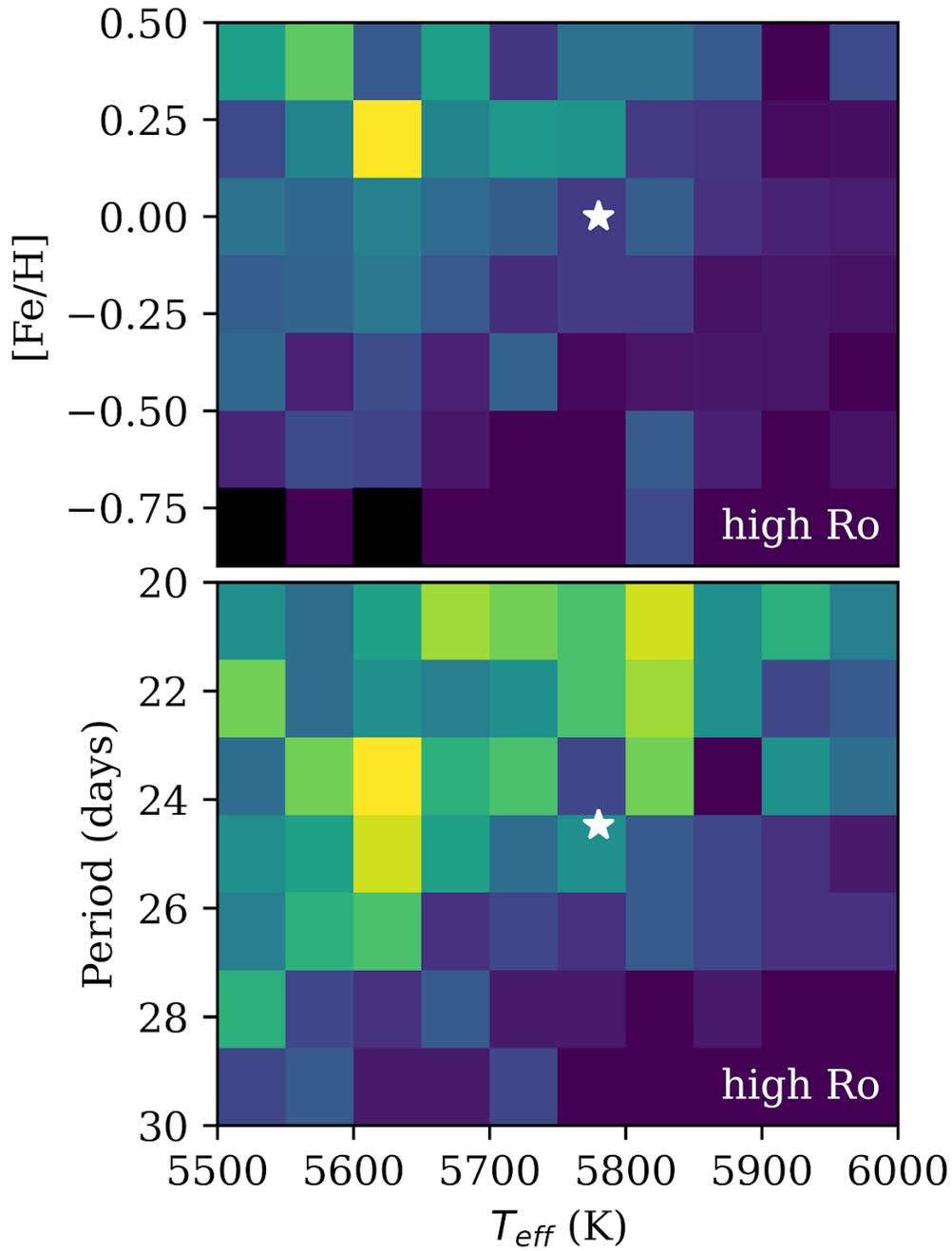

**Fig. 2. Fraction of stars with detected rotation (A) and distribution of detected rotation periods (B).** Rotation is detected in stars with surface temperatures and compositions that yield longer convective turnover times, and the detections are concentrated toward shorter periods. Both biases suggest lower values of *Ro* in the periodic sample. The Sun is indicated with a white star.